\documentclass[12pt,amsmath,amssymb,superscriptaddress]{revtex4}
\usepackage{graphicx}
\usepackage{dcolumn}
\usepackage{bm}
\begin{document}

\noindent \textbf{On the transient and steady state of mass-conserved reaction diffusion systems}

\vspace{5mm}

\noindent Shuji Ishihara$^{1}$, Mikiya Otsuji$^2$ and Atsushi Mochizuki$^1$\\

\vspace{3mm}

\noindent ${}^1$ Division of Theoretical Biology, National Institute for Basic Biology, 38 Nishigonaka, Myodaiji, Okazaki, Aichi, 444-8585, Japan\\
\noindent ${}^2$ Department of Anesthesiology, Faculty of Medicine, University of Tokyo, 7-3-1 Hongo, Bunkyo-ku, Tokyo 113-8655, Japan

\vspace{3mm}
\noindent E-mail: ishihara@nibb.ac.jp
%\begin{abstract}

\vspace{10mm}

\noindent \textbf{Abstract}: Reaction diffusion systems with Turing instability and mass
conservation are studied.  In such systems, abrupt decays of stripes
follow quasi-stationary states in sequence.  
At steady state, the distance between stripes is much longer than that estimated  by linear stability
analysis at a homogeneous state given by alternative stability conditions.  
We show that there exist
systems in which a one-stripe pattern is solely steady state for
an arbitrary size of the systems. The applicability to cell biology is discussed.
%\end{abstract}

%\pacs{82.40.Ck, 87.10.+e}

\baselineskip 24pt

\newpage

Pattern formation, the emergence of a spatial structure from an initially
uniform state, has been often studied on the framework of reaction
diffusion systems (RDS). It is extensively applied to physical,
chemical, and biological systems to explain their specific spatial
structures \cite{NicolisPrigigine,Meinhardt,Murray}. Turing instability is the most prominent mechanism, forming spatially
periodic stripes \cite{Turing}. The intrinsic distance between stripes is, in principle, estimated by
the linear stability analysis at a homogeneous state
\cite{NicolisPrigigine,Meinhardt,Murray,Turing}. However, this
estimation could be invalid when applied far from a uniform state. For example, a
second bifurcation can arise which would indicate the collapse of a simple periodic
structure \cite{ColletIoos}. In such situations, the transient
dynamics of pattern formation would be difficult to predict. In general,
RDS shows various dynamics even when steady states are reached
\cite{Nishiura}. So far, few studies have discussed transient
dynamics using the computational analysis of the famous Gray-Scott model
\cite{Nishiura,GrayScott} and by reduced dynamics on the slow
manifold \cite{KawasakiOhta,Ei2002,Kolokolnikov}.

In this Letter, we study a class of RDS in the context of the above aspects. We consider
RDS showing Turing instability, in which no production and no
degradation of substances occur \cite{IshiharaKaneko}. As we will see, the following
properties are observed in common; (i) The transient dynamics is a sequential
transition among quasi-steady states, with a decrease in the number of
stripes. (ii) The distance between resultant stripes cannot be estimated
from the linear analysis at uniform state. In particular, there are
systems in which a one-stripe pattern is solely stable state regardless
of the system size.

Consider a diffusible chemical component with two internal states, U and V.
Diffusion coefficients are $D_u$ and $D_v$, respectively, for
which we can set $D_u <D_v$ without loss of generality. The
transition rates between U and V are regulated by each other. We
studied a $1$-dimensional system with size $L (0\le x \le L)$ under
periodic boundary conditions unless otherwise stated. Concentrations in U and
V at position $x$ and at time $t$ are represented by $u(x,t)$ and
$v(x,t)$ respectively, and obey the following equations.
\begin{eqnarray}
 \partial_t u = D_u \partial_x^2 u-f(u,v) &  \label{eq:RDu} \\
 \partial_t v = D_v \partial_x^2 v+f(u,v) &  \label{eq:RDv} 
\end{eqnarray}
Obviously, the total quantity of the substances (total mass) is conserved.
\begin{equation}
s=\frac{1}{L}\int_0^L (u+v)dx. \label{eq:conserve}
\end{equation}
$s$ is the average concentration of the substance, which is determined
by the initial condition $u(x,0)$ and $v(x,0)$.

Below, all numerical simulations were performed with
$f(u,v)=au/(b+u^2)-v$ where $s=2.0$, $a=1.0$, $b=0.1$, $D_u=0.02$, and
$D_v=1.0$. We observed qualitatively the
same phenomena in several mass-conserved models \cite{Otsuji}.

\begin{figure}[tbhp]
\centering
\includegraphics[width=.85\textwidth]{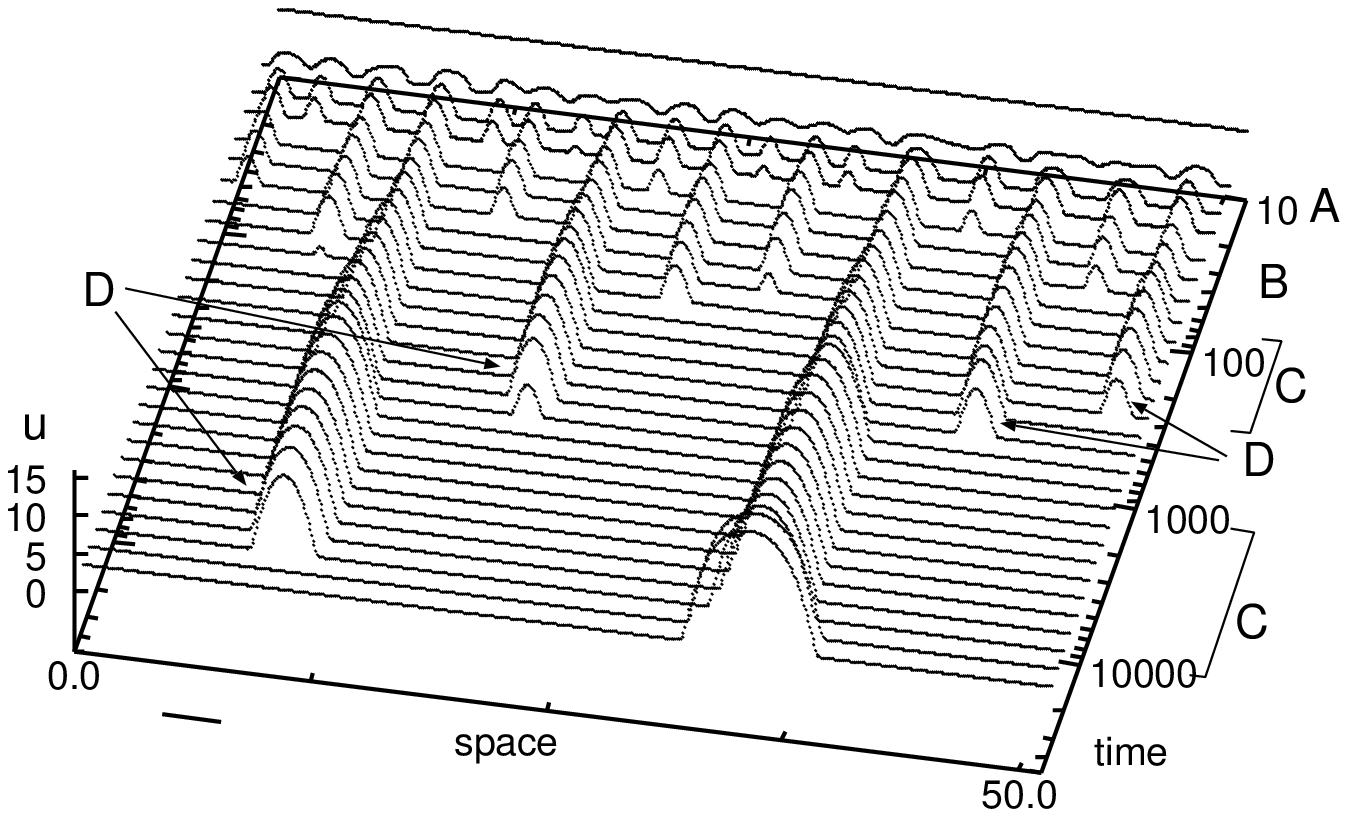}
\caption{ 
	Transient dynamics of a mass-conserved system. The model system and the meaning of 
	alphabet letters are explained in the text. System size was chosen as $L=50.0$ here. Note
	that time scale is represented by $\log$ scale. In the system, the most unstable wavelength at 
	homogeneous state is $\ell_{m}=3.2$, shown by the line segment in the left bottom. 
	We checked that the system eventually falls to a one-stripe pattern for any system size between 
	$2.0\le L \le 100.0$. 
	}
\label{fig:Dynamics}
\end{figure}

Uniform state $\vec{w}^*=(u^*,v^*)$ is derived from following the
conditions; $u^*+v^*=s$ and $f(u^*,v^*)=0$ (stable fixed point in kinetic
equation). Let $f_u^*~(f_v^*)$ be partial derivatives of $f$ with
regard to $u~(v)$ at $\vec{w}^*$. If the following relations are
satisfied, uniform state $\vec{w}^*$ loses its stability and the pattern
starts to rise.
\begin{eqnarray}
&f_v^*<f_u^*<0, &\label{eq:fvfu}\\
&D_u f_v^*-D_vf_u^* > 0. & \label{eq:DfuDfv}
\end{eqnarray}
All the waves ($e^{ikx} $) with wave number $k$ between $0< k^2 <
(D_uf_v^*-D_vf_u^*)/D_uD_v$ are unstable. At the beginning of the
dynamics, the wave with the largest instability grows as in usual Turing
systems (see A in Fig. \ref{fig:Dynamics}, and the line segment
representing the most unstable wavelength $\ell_m$).

In a mass-conserved system, characteristic transient processes
are observed. After the growth of a number of stripes (A in
Fig. \ref{fig:Dynamics}), some stripes stop growing and begin to decay
(B). With the decay of a stripe, neighboring stripes grow due to mass
conservation. The distance between neighboring stripes becomes larger
(B-C). If the distance is large enough the state appears to reach a steady
state (C, quasi-steady state). However, one (or more) stripe(s)
collapses abruptly with the concomitant growth of adjacent stripes (D). As the process
continues, the number of stripes decreases and the intervals between
the abrupt transitions gets longer (notice the $\log$-scale
representation in Fig. \ref{fig:Dynamics}). In
Fig. \ref{fig:Dynamics}, the system finally reaches a one-peak state.
The wavelength is much larger than $\ell_m$. Similar processes were
observed in many mass-conserved systems.

To understand the observed transient processes, consider the
stationary patterns of the system with size $L$. A stationary pattern
$\vec{w}_0(x)=(u_0(x),v_0(x))$ is given by the solution of
Eq. (\ref{eq:RDu},\ref{eq:RDv}) with left hand sides replaced by $0$.
In a mass-conserved system, there is a family of stationary solutions
parameterized by $s$, represented by $\vec{w}_0(x;s)$ explicitly. A
function $h(x)$ and a value $P$ exist, such that $u_0=h(x)/D_u$ and
$v_0(x)=(-h(x)+P)/D_v$, and satisfy the following equations.
\begin{eqnarray}
&P=D_uu_0(x)+D_vv_0(x) & \label{eq:Puv}\\
&\frac{d^2 h(x)}{dx^2}= f\left( \frac{h}{D_u},\frac{-h+P}{D_v}\right) &\label{eq:heq}.
\end{eqnarray}
Notice $D_uu_0(x)+D_vv_0(x) $ is independent of $x$. $P$ is related to
$s$ by $P=D_vs-(D_v-D_u)\bar{h}/D_u$, in which
$\bar{h}=\frac{1}{L}\int^L_0h(x)dx$ is the average of $h(x)$.

Let us represent the linear operator at a stationary state $\vec{w}_0$ by
$\mathcal{L}$. There are two eigen functions belonging to the zero eigen
value ($0$-eigen functions);
$\partial_x\vec{w}_0=(\partial_xu_0,\partial_xv_0)$ and
$\partial_s \vec{w}_0=(\partial_su_0,\partial_sv_0)$. The
former function is derived from the fact that the arbitrary translation of
stationary state, $\vec{w}_0(x+\Lambda)$ is also stationary, while
the latter is from the conservation property of the mass.  Conjugate
operator $\mathcal{L}^*$ is defined as the transposed matrix of
$\mathcal{L}$. One of the $0$-eigen functions of this operator is
$\vec{\phi}(x)=(1,1)$. The inner product between $\vec{f}=(f^u,f^v)$ and $\vec{g}=(g^u,g^v)$ (each defined on $0\le x\le L$) is to be $\langle \vec{f},\vec{g}\rangle_L\equiv \frac{1}{L}\int_0^L
(f^ug^u+f^vg^v)dx$. Then, $\vec{\phi}$ is thought to be the conjugate
vector of $\partial_s \vec{w}_0$ because $\langle
\vec{\phi},\partial_s \vec{w}_0\rangle_L=1$ and $\langle
\vec{\phi},\partial_x \vec{w}_0\rangle_L=0$.

Now, to evaluate the stability of a stationary pattern, consider the
following situation. Take the one-stripe stationary state $\vec{w}_0$ in
the system with $\frac{L}{2}$ length, which takes the minimum $u_0$ at
$x=0(=\frac{L}{2})$ and the maximum at $x=\frac{L}{4}$. Then copy
the exact same state on $\frac{L}{2}<x<L$, and name the system on $0 \le x\le L$
as the non-perturbed system (NPS). 
Left and right halves are independent of each other.
Next, the boundary
condition in NPS is changed at $x=0(=\frac{L}{2})$ and $\frac{L}{2}(=L)$ into
the usual periodic boundary condition of the system on $0\le x\le L$.
We refer to this modified system, which is the one we are interested in,
as the perturbed system (PS). We represent the state constructed as above
by $\vec{w}_0 \oplus \vec{w}_0$, where the left (right) hand side of
$\oplus$ represents the function on $0<x<\frac{L}{2}$
($\frac{L}{2}<x<L$). This state is obviously a stationary solution in
both NPS and PS.

Linear operators at the state are given by $\mathcal{L}_0$ for NPS and
$\mathcal{L}$ for PS. Because NPS is simply the juxtaposition of identical
systems, $\vec{\psi}^0_1=\partial_x \vec{w}_0\oplus \partial_x
\vec{w}_0$ and $\vec{\psi}^0_2=\partial_s \vec{w}_0\oplus \partial_s
\vec{w}_0$ are $0$-eigen functions for $\mathcal{L}_0$. They are also $0$-eigen functions in PS;
$\mathcal{L}_0\vec{\psi}^0_i=\mathcal{L}\vec{\psi}^0_i=0$ ($i=1, 2$).
Another $0$-eigen functions of $\mathcal{L}_0$ is
$\vec{\psi}^0_3=\partial_s \vec{w}_0\oplus (-\partial_s \vec{w}_0)$
but this is not $0$-eigen functions of $\mathcal{L}$ anymore
\cite{Note1}. However, when the amplitudes of $|\partial_s u_0|$ and
$|\partial_s v_0|$ are small at $x=0$ and $\frac{L}{2}$, the
discrepancy between NPS and PS is small and we can expect a
function $\vec{\psi}$ and a value $\lambda$ which are close to
$\vec{\psi}_3^0$ and $0$, respectively, and that they satisfy following relation.
\begin{equation}
\mathcal{L}\vec{\psi}=\lambda \vec{\psi} \label{eq:eigen-equation}
\end{equation}
If $\lambda$ is positive, then the stationary state $\vec{w}_0 \oplus
\vec{w}_0$ is unstable and small fluctuations grow as $\sim e^{\lambda
t}\vec{\psi}$. Because $\vec{\psi}$ is similar to
$\vec{\psi}_3^0=\partial_s \vec{w}_0\oplus (-\partial_s \vec{w}_0)$,
the corresponding dynamics appears as the decay of a stripe and the growth of the other, as is observed in the numerical simulations.  Note the
conjugate function of $\vec{\psi}_3^0$ is $\vec{\phi}_L=\vec{\phi}
\oplus (-\vec{\phi})$.

\begin{figure}[tbhp]
\centering
\includegraphics[width=.85\textwidth]{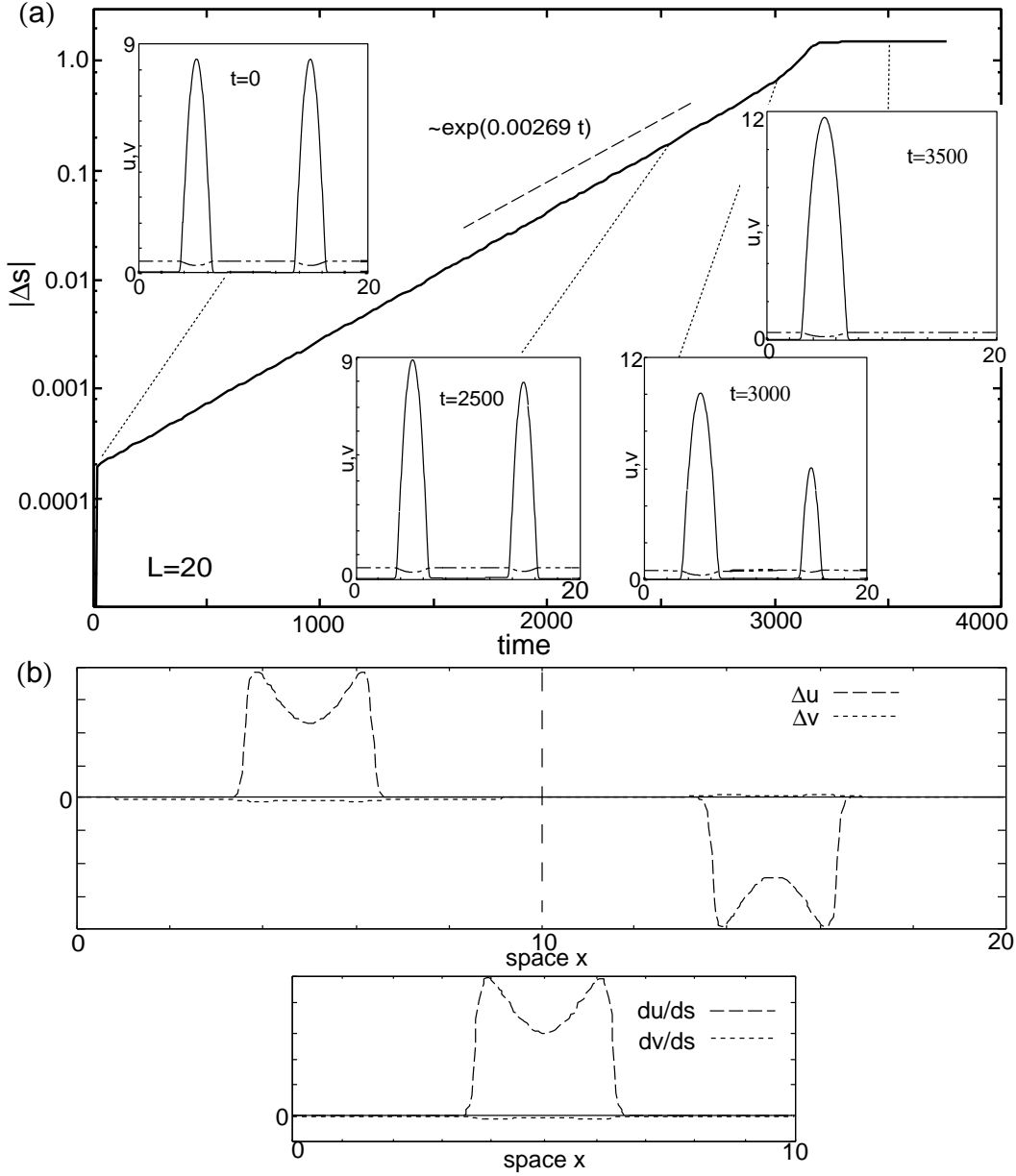}
\caption{ 
	Two identical steady solutions for $\frac{L}{2}=10.0$ are connected and perturbed 
	($\pm 1.0\%$, keeping total mass quantity) at $t=0$. (a) $\left| \Delta s \right|$ 
	grows exponentially with time, 
	indicating that one stripe decays while the other grows. $u(x,t)$ (solid) and $v(x,t)$ 
	(broken) at $t=0, 2500, 3000$ and $3500$ are 
	shown in insets. (b) $\Delta \vec{w}(x)$, the difference of $\vec{w}(x,t)$ between $t=200$ 
	and $300$ is shown in the top panel, while $\partial_s \vec{w}_0$ is shown in the bottom 
	panel (normalization is applied). 
	}
\label{fig:Dts}
\end{figure}

To check the validity of the above considerations, we numerically measured
some related quantities. At first, we simulated the equations in the
$\frac{L}{2}$-length system and obtained a steady one-stripe state. Then,
we extended the system size twice and copied the steady state to the
region $\frac{L}{2}\le x\le L$. Next, perturbations were added keeping the
total mass conserved and observed the resulting dynamics. $|\Delta s(t)|\equiv
|\langle\vec{\phi}_L,\vec{w}\rangle_L|
=\frac{1}{L}\left|\int_0^{\frac{L}{2}}(u+v)dx-\int_{\frac{L}{2}}^{L}(u+v)dx\right|$
is plotted in Fig. \ref{fig:Dts}(a) which shows the exponential growth of
the perturbation. Then, we compared $\vec{\psi}_3^0$ with the growing part of
$(u,v)$. In Fig. \ref{fig:Dts}(b), $\Delta \vec{w}(x)=(\Delta u(x),
\Delta v(x))$ is shown, where $\Delta
\vec{w}(x)=\vec{w}(x,t_2)-\vec{w}(x,t_1)$ is the difference of $\vec{w}$
between two growing time points $t_1$ and $t_2$. $\Delta \vec{w}(x)$
is similar to $\vec{\psi}_3^0$ which validates the above
considerations.

The expected $\vec{\psi}$ is a continuous and smooth function on $0\le x\le L$
and odd around $x=\frac{L}{2}$. Thus, it is enough to consider a
nontrivial solution of Eq. (\ref{eq:eigen-equation}) on $0\le x\le
\frac{L}{2}$ with boundary condition
$\vec{\psi}(0)=\vec{\psi}(\frac{L}{2})=0$. We can limit our arguments
on $0\le x\le \frac{L}{2}$ in the following discussion. We properly redefine
$\mathcal{L}$ under this limitation.

To obtain $\vec{\psi}$ and $\lambda$ in Eq. (\ref{eq:eigen-equation}),
$\vec{\psi}=\partial_s \vec{w}_0+\vec{\eta}$ is defined. Here,
$\vec{\eta}=(\eta^u,\eta^v)$ is orthogonal to $\partial_s \vec{w}_0$,
i.e.  $\langle\vec{\phi},\vec{\eta}\rangle_{L/2}=0$. In the first
order of approximation, $\vec{\eta}$ satisfies the relation
$\mathcal{L}\vec{\eta}=\lambda \partial_s\vec{w}_0$. The terms $m(x)$
and $Q^1(x)$ are introduced in $\eta^u(x)=m(x)/D_u$ and $\eta^v=(-m(x)+Q^1(x))/D_v$.
The following equations are then obeyed.
\begin{eqnarray}
& \frac{d^2 Q^1}{dx^2} = \lambda \left( \partial_s u_0+\partial_s
v_0\right) & \label{eq:dQ}\\ & \frac{d^2 m}{dx^2} -\left(
\frac{f_u}{D_u}-\frac{f_v}{D_v}\right)m=\frac{f_v}{D_v}Q^1(x)+\lambda \partial_s
u_0& \label{eq:dM}
\end{eqnarray}
From Eq. (\ref{eq:dQ}) and the boundary conditions, $Q(x) \equiv
D_u\partial_s u_0+D_v\partial_s v_0 + Q^1(x)$ is given by
$Q(x)=\lambda \hat{Q}(x)$ where
\begin{eqnarray}
\hat{Q}(x)=   \int ^x_0 \!\!\!dx' \!\!\int_0^{x'} \!\!\!\!\!dx'' \!\left( \partial_s u_0(x'')+\partial_s v_0(x'')\right)-\frac{L}{4}x \label{eq:Q}
\end{eqnarray}
By substituting $Q^1(x)$ into Eq. (\ref{eq:dM}), $m(x)=\lambda
A(x)-\partial_sP$ is obtained, where $A(x)$ is the solution of the
following equation with the boundary condition $A(0)=A(\frac{L}{2})=0$.
\begin{equation}
 \frac{d^2 A}{dx^2}-\left( \frac{f_u}{D_u}-\frac{f_v}{D_v}\right)A =
 \frac{f_v}{D_v} \hat{Q}(x) + \partial_s u_0
\end{equation}
 Due to the orthogonality
 $\langle\vec{\phi},\vec{\eta}\rangle_{L/2}=0$, $\lambda$ is given by
\begin{equation}
\lambda=\left( \frac{D_v-D_u}{D_uD_v}\bar{A} +\frac{1}{D_v} \bar{\hat{Q}}\right)^{-1} , \label{eq:lambda}
\end{equation}
where $\bar{A}=\frac{1}{L/2}\int_0^{\frac{L}{2}}A(x)dx$ and
$\bar{\hat{Q}}=\frac{1}{L/2}\int_0^{\frac{L}{2}}\hat{Q}(x)dx$.

\begin{figure}[tbhp]
\centering
\includegraphics[width=.85\textwidth]{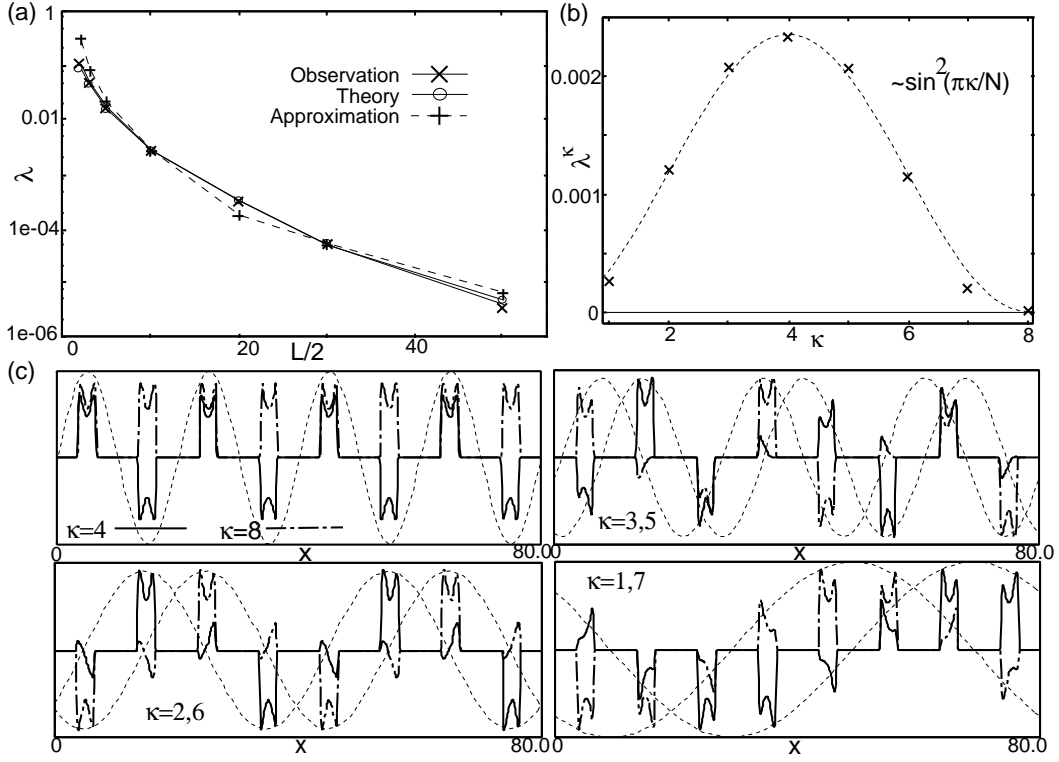}
\caption{ 
(a) Growth rates $\lambda$ for respective system sizes are evaluated from the observations of $|\Delta s(t)|$ ($\times$) and from Eq. (\ref{eq:lambda}) ($\bigcirc$). Approximated estimation of $\lambda$, $-\frac{4}{L^2}\partial_s P$, are also plotted ($+$). (b) Eigen values of $\mathcal{L}$ for the $8$-stripe state are numerically calculated ($L=80.0$), and (c) corresponding eigen functions. Note that eigen values except $\kappa=4, 8$ are degenerated. Sinusoidal curves are also shown for guidance.
	}
\label{fig:Epsilon}
\end{figure}

We calculated Eq. (\ref{eq:lambda}) numerically. Observed growth
rates and estimated values from Eq. (\ref{eq:lambda}) are plotted in
Fig. \ref{fig:Epsilon} (a) against half of the system size $\frac{L}{2}$,
which is the distance between two stripes. The two plots are in good agreement
with each other and support the validity of the arguments presented.

From $Q(x)=\lambda \hat{Q}(x)$, we can obtain $\lambda =
-\frac{4}{L}\frac{dQ(0)}{dx}$.  In NPS, $Q(x)$ is given by $\partial_s
P \oplus (-\partial_s P)$ and independent of $x$, while in PS $Q(x)$ is
connected with $0$ at $x=0$ and $=\frac{L}{2}$ by the modification of
boundary conditions. Thus, the sign of $\frac{dQ(0)}{dx}$ is the
same as $\partial_s P$ and the sign of $\lambda$ is the opposite of
$\partial_s P$. This leads to the condition of two
stripe instability in $L$-length system as
\begin{equation}
\partial_s P < 0. \label{eq:Pcond}
\end{equation}

The above arguments with the two stripes situation are extensible to an identical
$N$-stripe pattern, where each stripe has $\frac{L}{N}$ width.
Consider a set of independent functions
$\vec{\Psi}^k_0=\bigoplus_{j=1}^N e^{i\frac{2\pi}{N}\kappa j} \partial_s
\vec{w}_0$ $(\kappa=1, 2,\cdots,N)$, where $\bigoplus_{j=1}^N$ is defined
similarly to $\oplus$ and $ \vec{w}_0$ is redefined by the one-stripe
solution of $\frac{L}{N}$ width. Then eigen functions of
$\mathcal{L}$ (redefined for the $N$-stripe solution), $\vec{\Psi}^{\kappa}$, are
close to $\vec{\Psi}^{\kappa}_0$. Smooth connection of
$\vec{\Psi}^{\kappa}$ at each boundary $x=\frac{L}{N}j$ is conditioned.
Consideration of $Q(x)=D_u\Psi^{\kappa}_u+D_v \Psi^{\kappa}_v$, which is close to
$\bigoplus_{j=1}^N e^{i\frac{2\pi}{N}\kappa j}\partial_s P$, gives a rough
estimation of the eigen values as $\lambda^{\kappa} \sim
-4\left(\frac{N}{L}\right)^2 \partial_s P \sin^2 \left(\frac{\pi
\kappa}{N}\right)$. This indicates that a shorter wave (i.e. closer $\kappa$ to
$\frac{N}{2}$) has larger instability if $\partial_s P <0$.
$\lambda^{\frac{N}{2}}$ for even $N$ is identical to that estimated
from two stripes. Approximated values are shown in Fig. \ref{fig:Epsilon}.
The approximated $\lambda^1$ for $N=2$ is plotted in (a). Eigen values and corresponding eigen functions for $N=8, L=80.0$ are shown in (b, c).

To consider transient processes, an illustrative example can be seen 
from the stationary state of $2N$-stripes with a small perturbation. If
Eq. (\ref{eq:Pcond}) is satisfied for $\vec{w}_0$, the most
unstable function is $\Psi^{N}$. By the growth of the perturbation
along this function, the system reaches $N$-stripe pattern at last.  If this new state becomes unstable, a similar process follows
until the system reaches a steady state. In the (unstable) stationary state where the dynamics become close in their transient, $\lambda$ is small if the distance between adjacent stripes is
large. The corresponding state lasts for the duration of approximately $\lambda^{-1}$
and therefore each state appears quasi-stationary. Because the
distance becomes twice as large after each transient, the staying time in
the quasi-stationary state also gets longer. We could numerically
observe these processes from the $8$-stripe initial
condition. This demonstrates the underlying processes of the characteristic
transients inherent to mass-conserved systems.

After the long transient, the system reaches the steady state at which
the condition Eq. (\ref{eq:Pcond}) is violated. Thus, the
characteristic wavelength of the steady state is much longer than
expected by linear stability analysis at a uniform state in
a mass-conserved system. Our numerical model showed a one-stripe
solution eventually between $2.0 \le L\le 100.0$, while $\ell_m=3.2$
(data not shown).

Notice that Eq. (\ref{eq:DfuDfv}), the condition for Turing
instability, implies $\partial_s P <0$ at the uniform state.
Therefore Eq. (\ref{eq:Pcond}) is always satisfied in the early stages
of transient, where $\vec{w}(x)$ ranges in the neighborhood of
$\vec{w}^*$.

One interesting question that arises is the possibility of the system in
which the condition in Eq. (\ref{eq:Pcond}) is always valid in the
transient quasi-steady states except in sole stripe solutions.  Such
systems fall into a one-stripe solution after a long transient regardless
of system size. An example of such system was supplied by our group in \cite{Otsuji},
defined by $f(u,v)=-\alpha (u+v)\left((\delta u+v)(u+v)-\beta\right)$
where $\delta=D_u/D_v$. In this specific model, if $L$ is large
enough, $P$ is well approximated by
$P=\frac{3D_v\beta}{L\gamma}\frac{1}{s}$ with $\gamma \equiv\frac{1}{2}
\sqrt{\frac{D_v-D_u}{D_vD_u}\alpha\beta}$. Thus a one-stripe pattern is
the only stable state in the system. Though rigorous
conditions are not described here, many mass-conserved systems have such properties.

In this Letter, we study RDS in which the uniform state is
destabilized via a Turing mechanism and mass ($u+v$) is conserved.
The analysis presented is useful for stationary patterns in any RDS and
the conserved quantity does not have to be strictly defined by mass. Because the existence of any
conserved quantity brings the corresponding $0$-eigen function, our
arguments are applicable. Thus, the dynamics studied here may be
observed in a wider class of RDS with conserved quantities
\cite{AwazuKaneko,Narang}.

We did not mention the hierarchical structure of quasi-stationary
states in the phase-space, which is a necessary condition for the
sequential transient as discussed in \cite{Nishiura}. It is a global
property of the phase space of the systems and difficult to study.
Numerical simulations suggest it is satisfied in mass-conserved
systems.

Applications of this work are possible to many phenomena, particularly to
biological systems. Proposed biological models often contain conserved
quantities \cite{KeenerSneyd,Huang,Narang}. At the cellular level
($\sim 10 \mu m$), cytosolic proteins diffuse at $\sim 10 \mu
m^2/\mbox{sec}$ \cite{Elowitz} leading to the rough estimation of the
time scale of dynamics as $ \lambda^{-1}\sim \left(\Delta P/L^2\Delta s\right)^{-1} \sim L^2/D_v \sim 10 \mbox{sec}$. Typically, it is faster than the
synthesis or degradation of molecules and the dynamics is expected to occur
within the time scale in which mass-conserved modeling
is valid. The formation of cell polarity based on the above
discussions is a potential application \cite{Otsuji,Narang}.

S.I. is supported by Molecular-based New Computational Science Program.

\newpage

\newpage

\newpage

\end{document}